# Brain Network Construction and Classification Toolbox (BrainNetClass)

Zhen Zhou[a,b], Xiaobo Chen[c], Yu Zhang[d], Lishan Qiao[e], Renping Yu[f], Gang Pan[a,*], Han Zhang[b,*] and Dinggang Shen[b,g,*]

[a] College of Computer Science and Technology, Zhejiang University, Hangzhou 310027, China
[b] Department of Radiology and BRIC, University of North Carolina at Chapel Hill, Chapel Hill, NC 27599, USA
[c] Automotive Engineering Research Institute, Jiangsu University, Zhenjiang 212013, China
[d] Department of Psychiatry and Behavior Sciences, Stanford University, Stanford, CA 94305, USA
[e] School of Mathematics Science, Liaocheng University, Liaocheng 252000, China
[f] School of Electric Engineering, Zhengzhou University, Zhengzhou 450001, China
[g] Department of Brain and Cognitive Engineering, Korea University, Seoul 02841, Republic of Korea

## ABSTRACT

Brain functional network has become an increasingly used approach in understanding brain functions and diseases. Many network construction methods have been developed, whereas the majority of the studies still used static pairwise Pearson's correlation-based functional connectivity. The goal of this work is to introduce a toolbox namely "Brain Network Construction and Classification" (BrainNetClass) to the field to promote more advanced brain network construction methods. It comprises various brain network construction methods, including some state-of-the-art methods that were recently developed to capture more complex interactions among brain regions along with connectome feature extraction, reduction, parameter optimization towards network-based individualized classification. BrainNetClass is a MATLAB-based, open-source, cross-platform toolbox with graphical user-friendly interfaces for cognitive and clinical neuroscientists to perform rigorous computer-aided diagnosis with interpretable result presentations even though they do not possess neuroimage computing and machine learning knowledge. We demonstrate the implementations of this toolbox on real datasets. BrainNetClass (v1.0) can be downloaded from https://github.com/zzstefan/BrainNetClass.

## 1. Introduction

Functional connectivity (FC) based on resting-state functional MRI (RS-fMRI) is one of the major methods for brain functional studies. It describes the functional interactions among anatomically separated brain regions, often interpreted as information exchange and remote communication, or functional integration (Allen et al., 2014; Hutchison et al., 2013; Leonardi et al., 2013; Van Dijk et al., 2009; Thomas Yeo et al., 2011). Large-scale, whole-brain FC network is believed to be essential neural substrates for complex cognitive functions and can be modeled as a complex graph, where nodes represent brain regions and edges are inter-regional FC (Hallquist and Hillary, 2018; Sporns, 2010; Van Den Heuvel and Pol, 2010). The network topological structure can be analyzed based on graph theory (Bullmore and Bassett, 2011). During the past decades, we have witnessed a broad application of brain FC network-based disease studies (Badhwar et al., 2017; Fornito et al., 2015).

While many studies focused on the group-level differences in brain functional network between patients and healthy controls based on statistical inference, an emerging trend is to utilize machine learning techniques to learn diagnostic features from the brain networks to conduct individualized classification (Arbabshirani et al., 2017; Dubois and Adolphs, 2016; Rathore et al., 2017). Such a computer-aided diagnosis is more helpful for the clinicians to identify the diseased subject (Shin et al., 2016), personalized treatment planning (Gudayol-Ferré et al., 2015; Miao et al., 2017, 2018), or outcome prediction of medical treatment (Liu et al., 2018; Nie et al., 2019). From a methodological point of view, brain functional network-based classification is essentially a pattern recognition problem, where contributing features (e.g., FC links or network properties) can be jointly learned and weighted in a multivariate manner toward a classification goal. It does not only help to facilitate patient–control separation but also benefit imaging biomarker detection for better understanding the neuropathology of brain diseases (Yu et al., 2017; Zhang et al., 2019c).

With fast development in both brain network modeling (Calhoun et al., 2014; Smith et al., 2011; Zhang et al., 2017a) and machine learning methods (Bishop, 2006), the RS-fMRI-based clinical studies have been transforming from bench- to bed-side at an unprecedented speed (Cui and Gong, 2018; Lemm et al., 2011; Pereira et al., 2009; Shen et al., 2017). However, the broad interests are not accompanied by sufficient analytic tools for researchers from multiple disciplines to conduct brain network construction and network-based classification. Neuroscientists and clinicians with their respective domain knowledge in a pressing need of biomarker detection are not always equipped with the same amount of knowledge on imaging analysis, network construction, and machine learning. Without the help of rigorously designed toolbox, they could face problems such as double dipping (training and testing the classification model with the same data) (Kriegeskorte et al., 2009). For the classifier with freely estimable parameters, arbitrary parameter predefinition, or *ad hoc* parameter selection is not uncommon (Demirci et al., 2008). All these issues could harm the generalization ability of the diagnostic model, leading to degraded reproducibility and hindering clinical applications. A toolbox with standardized and rigorous classification framework is highly demanding (Cui and Gong, 2018).

---

*Corresponding authors
✉ gpan@zju.edu.cn (G. Pan); hanzhang@med.unc.edu (H. Zhang); dgshen@med.unc.edu (D. Shen)
ORCID(s): 0000-0002-4049-6181 (G. Pan); 0000-0002-2645-8810 (H. Zhang)

In this paper, we present a novel toolbox, namely Brain Network Construction and Classification toolbox (BrainNetClass v1.0). BrainNetClass is a user-friendly graphical-user-interface (GUI)-based Matlab toolbox designed to help neuroscientists, doctors, and researchers in other fields who have limited coding and machine learning knowledge easily and rigorously work on advanced brain functional connectomics construction and connectomics-based individualized disease diagnosis or other classification tasks. It avoids time-consuming, error-pruning, and complicated operations by providing users with an easy-to-use, automated toolbox, which can turn BOLD (blood oxygen level dependent) RS-fMRI time series into one of the various types of brain functional networks and generate a strictly-designed classifier for disease diagnosis, as well as produces comprehensive and interpretable results for model evaluation and better understanding of disease pathology. It is hoped that this toolbox could be of help in standardization the methodology and boosting reproducibility, generalizability, and interpretability of the network-based classification.

Compared to the existing toolboxes, BrainNetClass v1.0 has the following unique features: 1) It provides abundant algorithms for state-of-the-art brain network constructions, such as the recently developed high-order functional networks for measuring higher level of FC and representing more complex brain functional organization principals (Zhang et al., 2017a), as well as a battery of advanced sparse representation-based brain network construction methods that could generate more robust and biologically meaningful FC networks; 2) It provides an automated, standardized and well-accepted network-based classification framework with parameter optimization through nested cross-validation; 3) It offers a comprehensive battery of result evaluation metrics, including a receiver operating characteristic (ROC) curve, parameter sensitivity test, model robustness test, a full log of model configuration and evaluation, discriminative features, among many others.

In the following sections, we introduce the brain functional network modeling methods in Section 2.1 and the classification procedure in Section 2.2, as well as the comprehensive result report in Section 2.3. After brief descriptions of the modules and a walk-through of the toolbox in Sections 3.1 and 3.2, we provided real applications on RS-fMRI data sets with different classification goals in Section 4. We finish up with several key discussions in Section 5, especially the practical guidance on how to choose the proper network modeling algorithm.

## 2. Materials and Methods

### 2.1. Functional network construction

Pearson's correlation (PC) analysis between BOLD RS-fMRI signals of any pair of ROIs is the most popular FC network construction method. PC is intuitive and easy to interpret, but only captures the pairwise relationship. To characterize multi-ROI relationship, partial correlation or, more generally, sparse representation (SR) has been proposed for network construction, where the BOLD signal of a brain region is represented by a linear combination of the signals from other regions. Both types of methods have dominated the network analysis of RS-fMRI. However, each of them has shortcomings. PC only measures first-order collinearity between BOLD signals, but the pairwise relationship between two ROIs is not such simple. Among many recent research trends such as dynamic FC (Calhoun et al., 2014), one of the recent directions is to define "high-order" FC (HOFC) that is not measured based on "low-level" features (BOLD signals), which provides complementary information of higher-level relationship between two brain regions (Zhang et al., 2016a, 2017a, 2019b). On the other hand, SR usually results in sparse brain network because the added sparsity regularization term shrinks many weak FC links to be zeros. This makes the SR-based network less biologically meaningful. Recent research works have been focusing on building a biologically meaningful SR-based network by changing the existing or adding the new regularization terms (Yu et al., 2017; Zhang et al., 2019c).

The toolbox reflects the aforementioned research trends by integrating these state-of-the-art network construction methods, together with the traditional ones (PC and SR). In the next paragraphs, we will provide a brief introduction of each network construction method included in the toolbox; for more details, please refer to the original papers. For better organization, these algorithms are categorized into two types, 1) pairwise and 2) multi-ROI-based (or SR-based) network construction. In the GUI, there are also two types of networks containing the same algorithms; to help general users to understand, they are separated into the methods without the need of parameter optimization, and the methods requiring parameter optimization. Table 1 summarizes the meaning of the symbols used later.

#### 2.1.1. Pairwise FC-based network construction methods

Given a brain parcellation atlas with $N$ ROIs, RS-fMRI signal at the $i^{th}$ ROI can be represented as a column vector $\mathbf{x}_i = [x_{1i}, x_{2i}, ..., x_{Ti}]' \in R^T$ (where $'$ denotes matrix transpose). All the $\mathbf{x}_i$ ($i = 1, 2, ..., N$) results in a data matrix $\mathbf{X} = [\mathbf{x}_1, \mathbf{x}_2, ..., \mathbf{x}_N] \in R^{T \times N}$. The PC-based brain functional network can be represented as a graph with an edge weight matrix $\mathbf{W} \in R^{N \times N}$ with its element $w$ representing PC-based FC, characterizing simple pairwise temporal correlation of the raw BOLD signals. The PC-derived FC network usually serves as a baseline network construction method to compare with other advanced network construction methods (as such, the PC-based FC network is also referred to as low-order FC (LOFC), in contrast to the high-order FC methods (HOFC), as defined below).

**Table 1**
Symbols used in the current study

| Symbol | Meaning |
| --- | --- |
| L | Window length for sliding windows for dynamic FC |
| M | Number of subjects |
| N | Number of brain ROIs |
| T | Number of time points in RS-fMRI data |
| K | Number of clusters (new nodes) for dHOFC network |
| **W** | Connectivity matrix, or brain functional network |
| $w_{ij}$ | FC weight of an edge connecting two nodes $(i, j)$ in a network **W** |
| **X**, **X**$^m$ | RS-fMRI data matrix of the $m^{th}$ subject |
| **x**$_i$ | Mean RS-fMRI time series of the $i^{th}$ ROI |

In a similar manner, with each ROI's topographical FC profiles used as high-level features (instead of raw BOLD signals) in a PC analysis, we generate **topographical profile similarity-based HOFC (tHOFC)** between each pair of ROIs. Each $tHOFC_{ij}$ can be calculated by:

$$tHOFC_{ij} = \frac{\sum_k (w_{ik} - \overline{\mathbf{w_{i.}}})(w_{jk} - \overline{\mathbf{w_{j.}}})}{\sqrt{\sum_k (w_{ik} - \overline{\mathbf{w_{i.}}})^2}\sqrt{\sum_k (w_{jk} - \overline{\mathbf{w_{j.}}})^2}} \quad (1)$$

where $\mathbf{w}_{i.} = \{w_{ik}|k \in N, k \neq i\}$ and $i, j, k = 1, 2, ..., N, k \neq i, j$. It can be seen that it is the LOFC that is used in the PC calculation, which makes the resultant tHOFC different from LOFC between the same pair of ROIs. It has been shown that tHOFC could provide supplementary information to the conventional LOFC and help revealing additional group differences between Mild Cognitive Impairment (MCI) subjects and cognitively normal controls (Zhang et al., 2016a, 2019a).

The associated HOFC (aHOFC) is further defined based on the pairwise PC between the topographical profiles of tHOFC and those of LOFC for any pair of ROIs, in a similar manner as PC and tHOFC (Eq. 2). It measures the interlevel (between low-level and high-level) functional associations and complements the LOFC and tHOFC. Including all three types of the pairwise functional association indices (namely hybrid HOFC) could further improve the diagnostic accuracy of MCI (Zhang et al., 2017b).

$$aHOFC_{ij} = \frac{\sum_k (tHOFC_{ik} - \overline{\mathbf{tHOFC_{i.}}})(w_{jk} - \overline{\mathbf{w_{j.}}})}{\sqrt{\sum_k (tHOFC_{ik} - \overline{\mathbf{tHOFC_{i.}}})^2}\sqrt{\sum_k (w_{jk} - \overline{\mathbf{w_{j.}}})^2}} \quad (2)$$

Increasing evidence has shown that the FC calculated from ~1-min RS-fMRI data is varying across time and such variation could not be pure noise but reflect brain flexibility and moment-to-moment adaption (Gonzalez-Castillo et al., 2015). Chen et al. (2016) proposed a new brain network construction method based on dynamic FC, namely dynamics-based HOFC (dHOFC). First, dynamic FC $w_{ij}(\theta)$ is first calculated between ROIs $i$ and $j$ based on BOLD RS-fMRI signals in the sliding windows ($\theta = 1, 2, ..., \Theta$), each of which includes a small temporal segment of RS-fMRI signals in a length of L with a step size $s$ (thus, the number of sliding windows $\Theta = \lfloor (T - L)/s \rfloor + 1$). Then, dHOFC is calculated based on the PC between any pair of dynamic FC time series (Eq. 3).

$$dHOFC_{ij,pq} = \frac{\sum_{\theta=1}^{\Theta} (w_{ij}(\theta) - \overline{w_{ij}(\cdot)})(w_{pq}(\theta) - \overline{w_{pq}(\cdot)})}{\sqrt{\sum_{\theta=1}^{\Theta} (w_{ij}(\theta) - \overline{w_{ij}(\cdot)})^2}\sqrt{\sum_{\theta=1}^{\Theta} (w_{pq}(\theta) - \overline{w_{pq}(\cdot)})^2}} \quad (3)$$

By definition, $dHOFC_{ij,pq}$ characterizes temporal synchronization of dynamic FC time series, which actually defines a "high-order" connectivity among four instead of two ROIs. Therefore, the dHOFC network is defined in a $R^{N^2 \times N^2}$ space, instead of PC, tHOFC and aHOFC networks (all defined in a $R^{N \times N}$ space). To reduce the exponentially increased dimensionality for better classification, the third step of dHOFC is to run a clustering algorithm to reduce the dimension of dHOFC from $N^2 \times N^2$ to $K \times K$, where $K$ is the number of clusters, or "high-order nodes" in the dHOFC network. This step is to group the dynamic FC time series with similar temporal patterns and the cluster averaged dynamic FC time series are used to construct a lower-dimension representation of the dHOFC $\in R^{K \times K}$. The window length L and the cluster number K are two important parameters for dHOFC.

### 2.1.2. Multi-region-based network construction methods

The PC-based fully connected network is computationally easy but could result in spurious connections. SR is thus proposed to address this problem by adding a $l_1$-norm regularized item in an optimization problem to find the network

**W** by minimizing the loss function mathematically denoted in Eq. 4.

$$\min_{\mathbf{W}} \frac{1}{2}\|\mathbf{X}-\mathbf{XW}\|_F^2 + \lambda\|\mathbf{W}\|_1 \qquad (4)$$

$\lambda > 0$ is a parameter controlling the network sparsity. A higher $\lambda$ forces more elements in **W** to be zeros (no connection). In the toolbox, for all the SR-based methods (including SR), an additional step is conducted to make the network symmetric via $\mathbf{W} \leftarrow (\mathbf{W} + \mathbf{W}')/2$ (Yu et al., 2017; Zhang et al., 2019c). SR serves as another baseline method in the toolbox. Later on, we introduce several advanced methods as variations to SR to make the resultant networks more biologically meaningful, e.g., preserve strong FC links that are shown in the PC network. Of note, BrainNetClass uses SLEP package v4.1[1] for optimization for all SR-related methods.

An FC-strength penalty was introduced in SR, namely **weighted sparse representation (WSR)** (Yu et al., 2017). In WSR, the sparse regularization is weighted and the weight $C_{ij} = exp(-FC_{ij}^2/\sigma)$ where $FC_{ij}$ is the PC-based FC strength between the $i^{th}$ ROI and the $j^{th}$ ROI and $\sigma$ is a positive parameter ($\sigma$ can be set as all subjects' mean of standard variation of absolute PC-based FC strengths) used to adjust the decay speed of the FC-based weights (Eq. 5). If the BOLD signals of two ROIs strongly synchronized (indicating the two ROIs could have strong FC), then their connection should be less penalized to make it more possible to be retained during the SR to preserve the biologically putative FC links. Vice versa, weak links should be penalized more and likely to be eliminated after the SR. It has been shown that the network constructed by WSR is more biologically meaningful and more suitable for disease diagnosis than SR (Yu et al., 2017).

$$\min_{\mathbf{W}} \frac{1}{2}\|\mathbf{X}-\mathbf{XW}\|_F^2 + \lambda_1\|\mathbf{C}\odot\mathbf{W}\|_1 \qquad (5)$$

In another SR-based method called **strength-weighted sparse group representation (WSGR)**, strong FC links can be well preserved as WSR, and the original structured FC information in the PC-derived network can be reflected as well, thanks to another regularization term encouraging a joint preservation or suppression of groups of FC links (Simon et al., 2013; Yu et al., 2017). To achieve this, the PC-derived FC links are first grouped into a few subsets $\{O_g,\ g = 1, 2, \ldots, (G \ll N)\}$, each of which has a predefined weight $d_g = exp(-E_g^2/\sigma)$, where $E_g$ is the averaged absolute PC-based FC strength for the subset $O_g$ and $\sigma$ is set as all subjects' mean of standard variation of absolute PC-based FC strengths, which is the same as that in the WSR method. Then, the WSGR can be formatted as Eq. 6, where $\|\mathbf{W}_{O_g}\|_2 = \sqrt{\sum_{(i,j)\in O_g}(w_{ij})^2}$ is a $l_2$-norm penalty for each subset $O_g$ for joint selection or de-selection. Collectively, WSGR results in an FC network featuring sparsity

[1] https://github.com/jiayuzhou/SLEP

(i.e., overall sparsity, controlled by $l_1$-norm penalty), connectivity, and group structure (i.e., group sparsity, controlled by $l_2$-norm penalty). WSGR has two parameters ($\lambda_1$ and $\lambda_2$) to optimize for balancing such a tradeoff.

$$\min_{\mathbf{W}} \frac{1}{2}\|\mathbf{X}-\mathbf{XW}\|_F^2 + \lambda_1\|\mathbf{C}\odot\mathbf{W}\|_1 + \lambda_2\sum_{g=1}^{G}d_g\|\mathbf{W}_{O_g}\|_2 \quad (6)$$

It is notable that the SR conducts a network for each subject independently. This could lead to large inter-subject variability in derived networks, which is unfavorable for subsequent classification, as it will increase within-group variability and make the between-group separation more difficult. **Group sparse representation (GSR)** is put forward to address this problem by jointly estimating non-zero connections across all subjects (Wee et al., 2014). It encourages the derived connectivity networks to have a similar topological structure across all the subjects through a $l_{2,1}$-norm regularizer, as formulated in Eq. (7), where $\mathbf{W} = [\mathbf{w}_i^1, \ldots, \mathbf{w}_i^m, \ldots, \mathbf{w}_i^M]$ denotes the regional one-to-all PC-derived FC profiles of the $i^{th}$ ROI for all M subjects and $\lambda$ controls the extent of group sparsity.

$$\min_{\mathbf{W}_i} \sum_{m=1}^{M}(\frac{1}{2}\|\mathbf{x}_i^m - \mathbf{X}_i^m\mathbf{w}_i^m\|_2^2) + \lambda\|\mathbf{W}_i\|_{2,1} \qquad (7)$$

Zhang et al. (2019) recently proposed another GSR, namely **strength and similarity guided GSR (SSGSR)**, to better facilitate subsequent group comparisons or classification. They assumed that the PC-derived FC networks should inherently have higher within-group similarity but lower between-group similarity (i.e., the network from a patient could be more similar than that from another patient but less similar to that from a healthy control). To this end, the inter-subject similarity of the PC-derived FC profiles can be used as a regularizer as the last term of Eq. 8. Let $\mathbf{w}_i^m$ and $\mathbf{w}_i^l$ be the regional one-to-all PC-derived FC profiles of the $i^{th}$ ROI from the $m^{th}$ subject and the $l^{th}$ subject, a graph Laplacian $\mathcal{L}_i$ can be obtained by $\mathcal{L}_i = \mathbf{D}_i - \mathbf{S}_i$, where $\mathbf{S}_i = [s_i^{m,l}] \in R^{N\times N}$ is a similarity matrix with element $s_i^{m,l} = exp(-\|w_{i.}^m - w_{i.}^l\|_2^2)$ measuring the FC-profile similarity for the $i^{th}$ ROI between the two subjects.

$$\min_{\mathbf{W}_i} \sum_{m=1}^{M}\left(\frac{1}{2}\|\mathbf{x}_i^m - \mathbf{X}_i^m\mathbf{w}_i^m\|_2^2\right) + \\ \lambda_1\|\mathbf{B}_i\odot\mathbf{W}_i\|_{2,1} + \lambda_2 tr(\mathbf{W}_i\mathcal{L}_i(\mathbf{W}_i)^T) \qquad (8)$$

The second term of Eq. 8 has a weight matrix $\mathbf{B}_i = [\mathbf{b}_i^1, \ldots, \mathbf{b}_i^m, \ldots, \mathbf{b}_i^M]$ with each column $\mathbf{b}_i^m$ characterizing the exponentially-transformed one-to-all PC-derived FC profiles $\{b_{i,j}^m = exp(-(w_{i,j}^m)^2)\}(i, j = 1, \ldots, N, i \neq j)$. This term is to penalize weak connectivity while still preserving group consistency of the estimated networks (i.e., weighted group sparsity). There are two parameters controlling the tradeoff between weighted group sparsity ($\lambda_1$) and between-subject variability ($\lambda_2$).

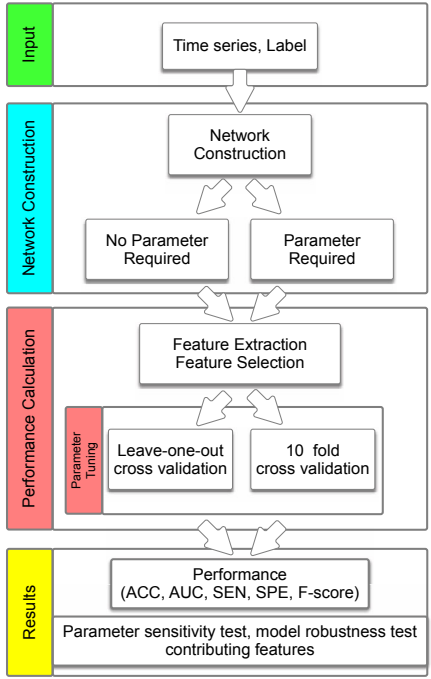

**Figure 1:** The Workflow of BrainNetClass Toolbox.

Qiao et al. (2016) proposed another functional brain network construction method, namely **sparse low-rank (SLR)** graph learning, by incorporating a low-rank prior into the SR-based network modeling. SLR results in a sparse yet modularity structure-preserved FC network, which is biologically meaningful considering the consensus finding of modular structure (more and stronger within-module connections, but less and weaker between-module connections) in the brain network (Bullmore and Sporns, 2012; Newman, 2006). It has been shown that, by increasing the modularity of the constructed FC network, the disease classification accuracy could be increased (Qiao et al., 2016). SLR is formulated in Eq. 9, where $\|\mathbf{W}\|_*$ is a trace norm (a.k.a. nuclear norm) that encourages the estimated adjacency matrix $\mathbf{W}$ to have a low-rank property, mathematically equal to larger modularity. It has two parameters ($\lambda_1$, controlling sparsity, and $\lambda_2$, controlling modularity) to be optimized.

$$\min_{\mathbf{W}} \|\mathbf{X} - \mathbf{XW}\|_F^2 + \lambda_1 \|\mathbf{W}\|_1 + \lambda_2 \|\mathbf{W}\|_* \qquad (9)$$

### 2.2. Network-based classification

After network being constructed, BrainNetClass (v1.0) continues to help conduct feature extraction from the constructed networks, feature selection (feature reduction), as well as training a classification model and testing it with cross-validation (Fig. 1).

While it is straightforward to treat each FC (or HOFC) link as feature (and this type of features are named as **connection coefficients**), there are also many studies using network properties calculated based on graph-theoretic analysis at the ROI-level as features, many of which have used clustering coefficients, one of the widely used regional characteristic calculated based on each ROI's local connectivity pattern (Rubinov and Sporns, 2010; Watts and Strogatz, 1998). Therefore, in addition to the FC coefficients, we also provide **local clustering coefficients** calculated based on weighted networks (Eq. 10) as another type of network features. For each network $\mathbf{W}$, weighted local clustering coefficient for the $i^{th}$ node can be calculated as:

$$f_i = \frac{2 \sum_{j: j \in \Omega_i} (w_{ij})^{\frac{1}{3}}}{|\Omega_i|(|\Omega_i| - 1)} \qquad (10)$$

where $\Omega_i$ is a set of nodes directly connected to the $i^{th}$ node and $|\Omega_i|$ denotes the number of elements in $\Omega_i$. The local clustering coefficients of each node in a weighted graph quantify the probability that the neighbors of this node are also connected to each other, which reflects local efficiency or "cliqueness" for each ROI in the network (Chen et al., 2016; Zhang et al., 2017b). With local clustering coefficients, the total number of features can be significantly reduced from $N \times N$ (or $K \times K$ for dHOFC) to $N$ (or $K$ for dHOFC). Of note, for certain network construction methods (e.g., dHOFC and SSGSR), we fixed the feature extraction methods to keep consistent to the previous studies (Chen et al., 2016; Zhang et al., 2019c) and to make as little choice as possible for users to reduce error. We also noted that there are many other network properties, such as shortest path length and betweenness centrality; and they could also be jointly used as features for better capturing the network topology (Liu et al., 2018).

The toolbox provides three widely used feature selection methods: two-sample $t$-test (Yu et al., 2016), least absolute shrinkage and selection operator (LASSO) (Tibshirani, 1996), and both (Zhang et al., 2019c), in order to further reduce redundant features, train a classifier with discriminative features only, and improve model generalizability with reduced feature dimension. If choosing two-sample $t$-test, each feature will be compared between two groups, and the ones with potential group difference ($p < 0.05$) will be used for training the classifier. LASSO is a multivariate feature selection algorithm that selects a few (sparse, controlled by a hyper-parameter $\lambda = 0.1$, which is fixed currently but users are able to change it if the number of selected features is too small or too many) features that are jointly informative to the classification. The two methods can be jointly used, so that the $t$-test roughly filters features and LASSO performs finer filtering. For certain methods, we fixed the feature extraction and selection to keep consistency with previous studies and to reduce unnecessary decision making.

We use a support vector machine (SVM) as the classifier in the toolbox. SVM is originally a mathematical approach based on the nonlinear optimization problem (Cortes

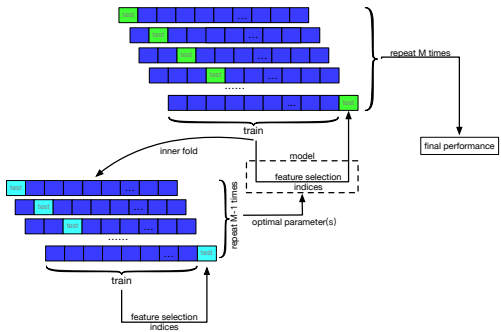

**Figure 2:** Nested LOOCV for parameter optimization.

**Table 2**
Definition of the classification evaluation metrics

| Measurement | Definition |
|---|---|
| ACC | $\frac{TP+TN}{TP+FP+TN+FN}$ |
| SEN(recall) | $\frac{TP}{TP+FN}$ |
| SPE | $\frac{TN}{TN+FP}$ |
| Youden | SEN+SPE-1 |
| BAC | $\frac{SPE+SEN}{2}$ |
| Precision | $\frac{TP}{TP+FP}$ |
| F-score | $2 \times \frac{precision \times recall}{precision+recall}$ |

TP, true positive; TN, true negative;

FP, false positive; FN, false negative.

and Vapnik, 1995). It is a popular method in many fields including the machine learning field (Misaki et al., 2010). It incorporates the concept of structural risk minimization by creating a separating hyperplane, which not only can maximize the margin separating two classes of data, but also can minimize the misclassification error. In this toolbox, we use the LIBSVM v3.23 (Chang and Lin, 2011) to perform classification based on SVM. Currently, the hyper-parameter $C$ of SVM is set to 1, but users are able to change it. Previous studies suggest that different machine learning algorithms may affect the model performance (Cui and Gong, 2018).

Cross-validation is important for machine learning but it could be done incorrectly for inexperienced users. Amongst many other strategies (Varoquaux et al., 2017), we provide two classic and popularly-used cross-validation strategies for users to choose: leave-one-out cross-validation (LOOCV, for small sample size) and 10-fold cross validation (for larger sample size). For LOOCV, a subject is treated as testing data and the other $M$ - 1 subjects are used for training the classifier. The above procedure is repeated for $M$ times, each time leaving out a different subject for testing. Finally, the performance is computed across $M$ classification results. In 10-fold cross-validation, all the $M$ subjects are evenly separated into 10 subsets, and the classifier training is carried out based on nine subsets and tested on the leftover one subset. Such a process is iterated until each of the 10 subsets has been used as the testing set. The whole process can be repeated for many times (default, 10 times), as the subject partitioning is random and the testing performance may heavily depend on the data partitioning. Finally, the classification performance is averaged across all 10 folds and all repetitions. Of note, the feature selection is also conducted inside the cross-validation (or nested cross-validation, see below), i.e., performing feature selection on the training data and applying the feature indices to the testing data.

For some network-based classification strategies, there are some freely estimable parameters to optimize (see Fig. 1 *Parameter-Required Network Construction*). To this end, in each iteration of the cross-validation, nested cross-validation will be implemented for parameter optimization (for LOOCV, the inner cross-validation is also LOOCV; for 10-fold cross-validation, the inner cross-validation is also 10-fold). For example, if LOOCV is chosen, from the $M$-

1 training subjects, one subject will be left out for validation and the rest $M$-2 are used for training with each combination of the parameters. This process is repeated for $M$-1 time and its classification accuracy under each specific parameter combination can be compared; the one leading to the best performance will be selected and used to construct the optimal classification model for testing with the outer LOOCV (Fig. 2).

Classification performance is evaluated based on a battery of assessment metrics, including classification accuracy (ACC), the area under ROC curve (AUC), sensitivity (SEN), specificity (SPE), precision, balance accuracy (BAC), Youden Index (Yonden), and F-score (Sokolova et al., 2006) (Table 2). Among them, the ROC curve is an important figure describing the diagnostic ability of a binary classifier when its discrimination threshold is varying. Its AUC measures the probability that a classifier assigns a higher score to a randomly chosen positive example than that to a randomly chosen negative example.

## 2.3. Result display and interpretation
### 2.3.1. Contributing features

In addition to the numeric classification performance evaluations and the ROC curve, the user may also want to know which features contribute more (a.k.a., contributing features) or are more important to the disease classification. We thus provide the averaged weights derived from the SVM for each feature across all cross-validation runs as feature importance measures. The larger the absolute value a feature has, the more important this feature could be. Therefore, it can be an $N \times N$ matrix if features are connection coefficients and a length-$N$ vector if features are local clustering coefficients. It is recommended to visualize the involved connectivity links or ROIs as the clinicians may think they are potential biomarkers of disease. For dHOFC, a total of $K$ matrices (each has a size of $N \times N$) will be also generated to identify the important "high-order" nodes (known as a cluster of synchronized dynamic FC links), which are essentially low-order networks, as shown in (Liu et al., 2018).

Another quantitative measurement of feature importance, i.e., the occurrence of each feature being selected in feature selection across all cross-validation runs, could also indicate feature importance (Chen et al., 2017a; Yu et al., 2017; Zhang et al., 2019c) and is also reported in the results. Similarly, it is an $N \times N$ matrix if features are connection coefficients and a length-$N$ (or length-$K$ for dHOFC) vector if features are local clustering coefficient. The more frequently a feature has been selected, the more important this feature could be.

### 2.3.2. Constructed networks

Aside from the discriminative features, the constructed brain functional networks are also informative to researchers conducting network-based classification. Therefore, we also show the group-averaged brain network in a form of weighted adjacency matrix for each group (or the averaged brain network constructed using the optimal parameter(s) if the user chooses one of the parameter-required brain network construction methods (Chen et al., 2017a)).

### 2.3.3. Log file

Meanwhile, a full log of model configuration and result report are summarized in a log file. The former includes the network construction method used, feature extraction and selection methods used, and the ranges of parameters to optimize, and the model evaluation (cross-validation) used. The latter summarizes all the numeric result, including all the model performance assessment metrics, as well as the suggested parameters (according to a parameter sensitivity test, see below), and the parameter selection occurrence, as reported by several studies (Chen et al., 2017a).

Additional unique features that BrainNetClass (v1.0) provides compared to other toolboxes are the suggested (model) parameters for future use according to a parameter sensitivity test and the model robustness test for calculating the most consistently chosen parameters, as separately explained in Sections 2.3.4-5.

### 2.3.4. Parameter sensitivity test and the suggested parameters

Most of the brain network construction methods provided in our toolbox require parameter tuning. It brings a fundamental question, i.e., for the future classification to the new subject, which parameters could be the best parameters to form the classification model. In addition, we would like to test if the achieved classification performance is very sensitive to specific parameter choices. To this end, our toolbox implements a comprehensive assessment of the relationship between model performance and parameters used, as did previously (Zhang et al., 2019c; Yu et al., 2017).

Specifically, the classification model is re-trained with each value (or each combination) of the freely estimable parameter(s) with LOOCV for performance evaluation. For the brain network construction with one parameter, this will create a bar plot showing the changing classification accuracies with different values of the parameter. For the brain network construction with two parameters, 2-D bar plots will be generated showing the accuracy changes at every combination of the two parameters. If the classification is sensitive to the parameters, there could be a bar that is significantly higher than every other bar, or a few bars are singled out as "islands" among other bars. Thus the model is not good enough, as it highly depends on a specific parameter or parameter combination. In this case, the model generalization ability might be low, and it is less likely the same model could work on other data from other research centers. On the other hand, if the contour spanned by the bars are *smoothly* changing from peak to low performance and the peak performance is quite similar to the performance reported by the rigorous nested cross-validation, it means the classification model is less sensitive to the parameters and could be trustful. In this case, the parameter(s) associated with the peak bar is (are) the *Suggested Parameters* for future use for the new coming samples. The suggested parameters will be reported in the main GUI and the log file, which indicates, in an *ideal* scenario (e.g., future usage), how the researcher should set up the model.

### 2.3.5. Model robustness test and the most consistently chosen parameters

In addition to the LOOCV-based parameter sensitivity test that focuses on the ideal performance with all possible parameters, model robustness test is to answer a question whether a classification model is robust so that every time in the nested cross-validation the same parameter(s) was (were) chosen as the optimized parameter(s). Therefore, how many times a specific parameter or a combination of parameters has been selected as the optimal parameter(s) is recorded by the toolbox (namely *Parameter Selection Occurrence*). If the parameter is selected significantly more than others, the classification model is robust. On the other hand, quite evenly distributed parameter selection occurrence indicates no dominant parameter (or parameter combination), which means that the model could drastically change even with only a few training data being changed (or less robust). Of note, it is not necessary the parameter(s) with the highest occurrence is the same as the suggested parameter(s), but we have observed that a good classification model has the suggested parameter(s) corresponding to the high occurrence.

## 3. Toolbox Design and Usage

### 3.1. Functional modules and designing logics

BrainNetClass (v1.0) consists of several sequentially executed modules, including brain network construction, feature extraction, feature selection, model evaluation (cross-validation), parameter sensitivity test, and result generation. The workflow of the toolbox is shown in Fig. 1 and can also be easily seen in the main GUI (Fig. 3(a)).

The preprocessed RS-fMRI time series serve as inputs into the toolbox. These regional averaged time series can be easily generated by other toolboxes, such as SPM[2] or FSL[3]

---

, and take a text format. Another input will be the labels for all subjects, which also takes a text format. The I/O is similar and intuitive, similar to many other toolboxes. As the input format follows the output format of other toolboxes like SPM and FSL, the user does not need to do any further preprocessing.

The user needs to decide which brain network construction method to use. All the network construction algorithms are organized into two types: Type I for those without any parameter to optimize (PC, aHOFC, and tHOFC) and Type II for those with parameter optimization required during the network construction (SR, GSR, WSR, WSGR, SGR, SS-GSR, SLR, and dHOFC). Most of the parameter-required network construction methods are SR-based. These methods gain certain advantages but need parameter optimization, which brings additional issues such as reduced training samples (as some of them are used for parameter optimization) and increased risk of overfitting (i.e., more works need to be done to test model robustness and parameter sensitivity). It is recommended that users select only one method that is the most appropriate for their own study (how to choose the proper network modeling algorithm will be provided in Section 5.1). This is to avoid the blind selection of all methods and only reporting the one with the best result (because this violates the rule of machine learning as it is incorrect to determine the model based on testing data).

If the user chooses a method without any parameter, he/she will also need to determine the methods for feature extraction (connection coefficients or local clustering coefficients) and feature selection (*t*-test, LASSO, or both). If the user chooses a method that needs parameter optimization, the feature extraction and selection will be automatically filled up and the default parameter range will be provided. All the default setting is to reduce unnecessary decision making and possible errors. The user can change the default parameter range based on their own preference. The user either chooses LOOCV or 10-fold cross-validation to evaluate the classification model depending on the number of sample size.

As the toolbox goes through every module, it generates multiple results and the classification performance is shown in the *result* window and the suggested parameter(s) is shown in the last window next to it; if applicable and the user chooses to perform the parameter sensitivity test. In the result folder specified by the user, there will be many other results and figures generated, together with a log file summarizing the details of the method used for assisting paper writing. It is advocated that not only the diagnosis accuracy but also the constructed networks, the contributing features, and the model robustness (i.e., whether the model tends to pick specific parameter(s) when the data is changed during cross-validations) should be recorded and reported in the network-based individualized classification papers. Collectively, the toolbox will guide users through all the setup in every module for network construction and classification, and show the final results in the GUI as the end of the process. All the model information can be retrieved even the user closes the GUI. The user is able to interpret the result according to their domain knowledge based on all the additional outputs provided by the toolbox.

## 3.2. Step-by-step usage

We give a brief step-by-step instruction of the toolbox usage (with more detailed demonstration included in the toolbox manual):

- Specify the RS-fMRI time series data of all subjects by selecting the folder containing all the text files (in each text file, the data is arranged as a matrix sized $T \times N$). Also, specify a text-formatted label file containing a column of labels for all subjects (e.g., -1 for patient and 1 for control) in the same order as what the Matlab takes when reading these time series data. The output directory should be also specified (Fig. 3(a)).

- Choose the network construction method by first selecting a type of brain construction methods and then the specific method. When choosing a parameter required brain network construction method, the user needs to specify the parameter range(s) if they do not want to use the default settings (Fig. 3(b)). There are brief explanations of the meanings of the parameters on the panel above parameter settings for users to check.

- Select or use predefined feature extraction and feature selection methods. There are also explanations on the panel above for users to check.

- Choose model evaluation or cross-validation method. If choosing 10-fold cross-validation, users might also want to specify how many times the 10-fold cross-validation will be repeated.

- After clicking the Run button and waiting for all the processes to complete, a "All Jobs Completed" window will pop out. Then all the results will be printed out on the result panel and the suggested parameters panel (if applicable), as shown in Fig.3(c).

- A full log of results for a hassle-free report is also generated in the result folder, as shown in Fig. 3(d).

- The users may also want to test the performance of other baseline methods, such as PC and SR, to compare with the state-of-the-art methods by repeating the above steps.

## 4. Toolbox Validation

To further evaluate our toolbox, we applied it to real RS-fMRI datasets. For each application, we chose one advanced method to construct brain network and compared the classification performance with the two baseline methods, i.e., PC and SR. Both the experiments were conducted using Matlab version 2018a based on a Window desktop computer with six CPU kernels and 64G physical memory. Before computing, the estimated required memory will be displayed to users for them to decide whether more powerful computational resources should be used. The memory required and the computing time are proportional to the number of ROIs and the sample size, and depend on the network construction method (generally, dHOFC consumes more memory, and SR-related methods with two estimable parameters need more computational time).

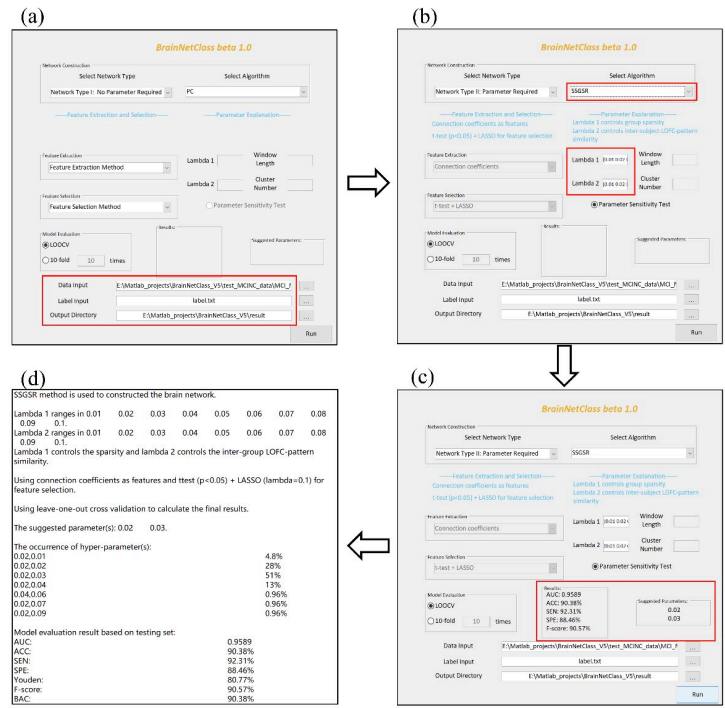

**Figure 3:** Step-by-step setup in BrianNetClass.

## 4.1. Application 1: SSGSR-based MCI diagnosis
### 4.1.1. Materials and methods

The dataset is from the Alzheimer's Disease Neuroimaging Initiative (ADNI)[4]. MCI is a transitional stage of brain cognitive decline between Alzheimer's Disease (AD) and cognitively normal people, which has high chance to progress to AD within 5 years (Gauthier et al., 2006). This application is to demonstrate the feasibility of timely detection of MCI based on brain functional networks.

The RS-fMRI data from 52 normal controls (NC) and 52 MCI patients were selected from the ADNI-2 dataset. The two groups are age- and gender-matched and they were all scanned using 3.0-T Philips scanners (see detailed imaging protocols elsewhere[5] ). The data were from multiple imaging centers but the quality control was carefully carried out to make sure of its inter-site consistency. The RS-fMRI data were preprocessed by using SPM8 with the procedure described elsewhere (Chen et al., 2017b). Automated Anatom-ical Labeling (AAL) template was used to extract the ROI time series from the 116 ROIs.

We use the SSGSR to construct brain networks for all the subjects. The parameters including $\lambda_1$ (controlling group sparsity) ranging [0.01, 0.02, ..., 0.1] and $\lambda_2$ (controlling inter-subject network similarity ranging [0.01, 0.02, ..., 0.1]. The connection coefficients were used as features. The two-sample $t$-test ($p < 0.05$, uncorrected) was adopted to initially reduce less discriminative features and LASSO was used to further select a subset of the features by removing the redundant ones. We compared the performance of the SSGSR model with PC and SR. LOOCV was adopted to evaluate the performance. The effectiveness of SSGSR method could be affected by $\lambda_1$ and $\lambda_2$. With the nested LOOCV being automatically executed, the optimal parameter values are determined based on the inner LOOCV.

### 4.1.2. Results

The SSGSR achieved much better performance than PC and SR (Table 3, Fig. 4(a)). As shown in Fig. 4(b), the model was more sensitive to $\lambda_1$ than $\lambda_2$. The suggested parameters are $\lambda_1 = 0.02$ and $\lambda_2 = 0.03$. The classification accuracy is 90.38% yielded with nested LOOCV (Table 3), which is close to the accuracy with the suggested parameters (93.27%). The most selected parameters (shown in Fig. 4(c), indicating that more than 50% of the times such a combination of the parameters was selected) are the same as the suggested parameters, indicating fair robustness of our model.

The connectivity features with potential prognostic values were plotted, showing the FC links that are 100% being selected during all the LOOCV runs (Fig. 4(d)), with the link thickness and the node size indicating the importance of the features. A total of 17 discriminative connections were identified. Most of these regions and connections have been shown closely related to AD pathology in previous studies (Buckner et al., 2008, 2005; Frisoni et al., 2009; Jacobs et al., 2012; Li et al., 2012; Thomann et al., 2008; Wee et al., 2016, 2012), such as the default mode network that plays an important role in high-level cognitive functions and the connections between cerebellum and cortical regions. Of note, all the results can be retrieved from the saved results, including the ROC curves, the suggested parameters, the model robustness testing result, and the most important features. The user can use visualization toolbox like BrainNet Viewer (Xia et al., 2013) to display them for helping interpretation.

---

[4] http://adni.loni.usc.edu
[5] http://adni.loni.usc.edu/methods/documents/mri-protocols

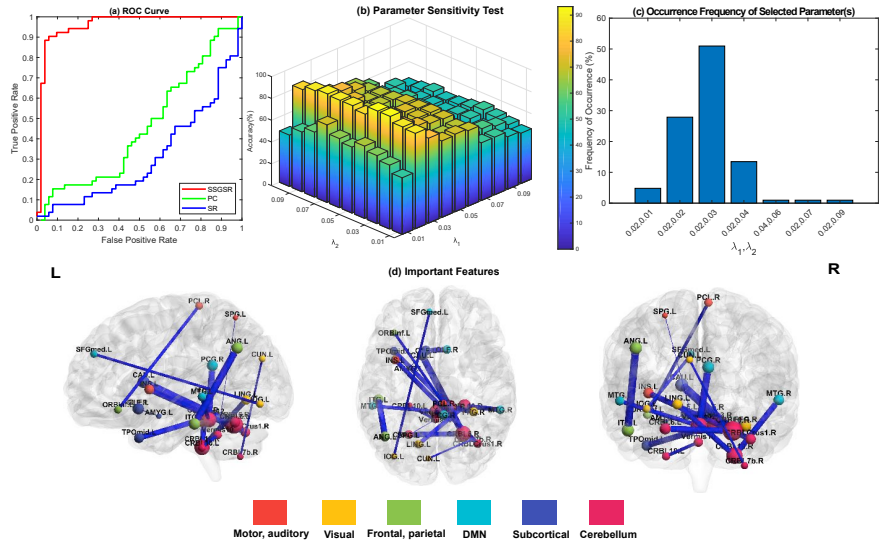

**Figure 4:** MCI diagnosis results, including the ROC curve comparisons between SSGSR and PC/SR (a), parameter sensitivity testing result with all possible combinations of the parameters (b), and the parameter selection occurrence for model robustness evaluation (c), and the contributing features suggested by the toolbox according to the occurrence of being selected in the cross-validations (d) plotted by using BrainNet Viewer, where the connections 100% selected were shown, with edge thickness representing the averaged absolute weight (indicating feature importance) and the node size representing the averaged absolute weights associated with each node. The suggested parameters are $\lambda_1 = 0.02$ and $\lambda_2 = 0.03$, based on which the model reaches the highest accuracy (93.27%) when evaluated based on LOOCV (b). The suggested parameters are also the most selected parameters (c).

**Table 3**
MCI diagnostic performance by using SSGSR, PC, and SR

|  | AUC | ACC | SPE | Youden | BAC | SEN | F-score |
|---|---|---|---|---|---|---|---|
| SSGSR | 0.9589 | 90.38% | 88.46% | 80.77% | 90.39% | 92.31% | 90.57% |
| PC | 0.4571 | 46.15% | 38.46% | -7.69% | 46.16% | 53.85% | 50.00% |
| SR | 0.3081 | 34.62% | 38.46% | -30.77% | 34.62% | 30.77% | 32.00% |

## 4.2. Application 2: SGR-based subject's state classification

### 4.2.1. Materials and methods

Eyes open (EC) and eyes closed (EC) resting states have both used in many RS-fMRI studies, but several studies have shown that there are fundamental differences between these two states (Liang et al., 2014; Yu-Feng et al., 2007; Zhou et al., 2018). In this study, we aimed to evaluate the feasibility of such state classification based on brain functional connectivity in an individualized manner. In addition to the previous region activity-based studies, we also aimed to find out important FC links that played significant roles in the classification between EC and EO.

The data from 48 (22 females) college students (aged 19-31 years) was downloaded from a publicly available dataset, Beijing Eyes Open Eyes Closed Study[6]. The RS-fMRI data during EC and EO were separately acquired from the same subject using a SIEMENS TRIO 3.0-T scanner at the Beijing Normal University and the imaging protocol can be found in (Liu et al., 2013). One subject was excluded due to incomplete RS-fMRI data. The conventional RS-fMRI preprocessing was conducted using DPABI (Yan et al., 2016). None of these subjects was excluded due to excessive (> 2 mm in displacement or > 2° in rotation, or with mean Framewise Displacement (FD) > 0.5 mm) head motion.

SGR was used to construct brain networks consisting of

---
[6] http://fcon_1000.projects.nitrc.org/indi/retro/BeijingEOEC.html

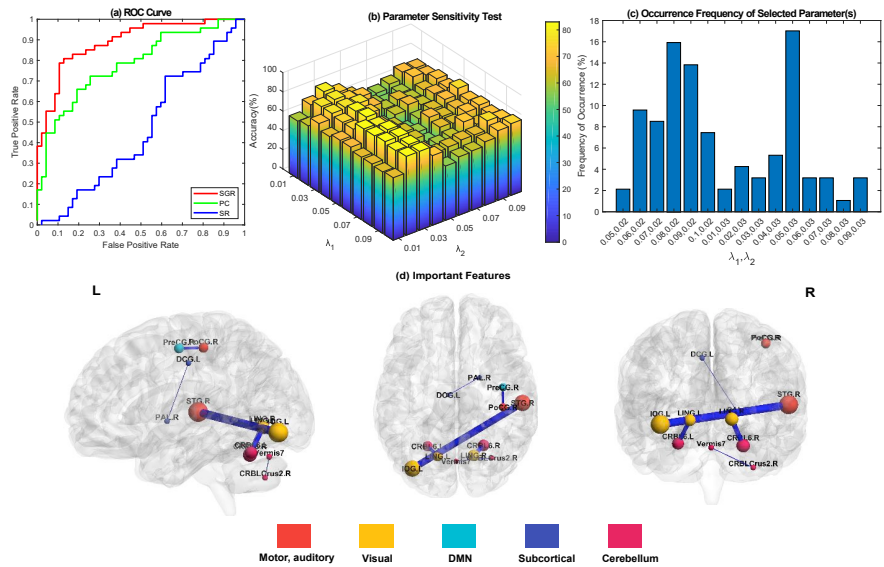

**Figure 5:** Eyes close vs. eyes open (EC vs. EO) classification results, including ROC curves for the SGR, PC, and SR (a), result from parameter sensitivity test (b), the parameter selection occurrence (c), and the important features identified (d).

**Table 4**
The classification performance of EC/EO by SGR, PC, and SR

|        | AUC    | ACC    | SPE    | Youden  | BAC    | SEN    | F-score |
|--------|--------|--------|--------|---------|--------|--------|---------|
| SSGSR  | 0.8927 | 79.79% | 76.60% | 59.98%  | 79.79% | 82.98% | 80.41%  |
| PC     | 0.7841 | 71.28% | 70.21% | 42.55%  | 71.28% | 72.34% | 71.58%  |
| SR     | 0.4595 | 43.62% | 44.68% | -12.77% | 43.62% | 42.55% | 43.01%  |

116 ROIs from the AAL template. The parameters of the SGR model were set up as follows: $\lambda_1 = [0.01, 0.02, ..., 0.1]$ and $\lambda_2 = [0.01, 0.02, ..., 0.1]$. All the other settings are the same as those in Application 1. The performance of the SGR model was compared with those of PC and SR based on LOOCV.

### 4.2.2. Results

The performance of SGR was better than that of PC and SR, but the method as simple as PC also resulted in satisfactory accuracy (Fig. 5(a), Table 4). The parameter sensitivity test shows that the results were not quite sensitive to the parameters (Fig. 5(b)), and suggested parameters ($\lambda_1 = 0.08$ and $\lambda_2 = 0.02$) resulted in similar accuracy as that derived from nested LOOCV (82.98% vs. 79.79%). However, the suggested parameters are not the most selected parameters (actually, they are the second most selected parameters, as shown in Fig. 5(c)). The mostly and second mostly selected parameters were only selected for 16 and 17 times (~34% and 36% across all the LOOCV runs), indicating that the model robustness should be further investigated and the result in Fig. 5(d) and Table 4 should be carefully interpreted because the performance and the identified contributing features from SGR were not derived from the model with the same parameters. In this case, the PC could be the most suitable model, as it involves no parameters. Nevertheless, we also plotted the six most consistently selected connections in Fig. 5(d). Most of the regions and connections found are consistent with previous studies, such as the sensorimotor and visual cortices (Zou et al., 2015; Zhou et al., 2018).

## 5. Discussion

In this paper, we present an easy-to-use Matlab toolbox (BrainNetClass v1.0) for brain network construction and classification, which integrates some state-of-the-art network construction methods and provides a comprehensive solution of individualized classification with scientific rigor. This toolbox is designed to encourage clinical applications of brain functional network based on resting-state fMRI. The target users are the clinicians with data and domain knowledge but not sure about machine learning. The toolbox provides a standard and widely-adopted pipeline for network-based classification to minimize possible confusion without compromising freedom such as parameter optimization. It generates multi-facet qualitative (e.g., network visualization) and quantitative (e.g., model performance, model robustness) results, and allows users to explore the contributing features to further push the boundary of machine learning studies in the clinical field. Effectiveness of the toolbox was proven by real RS-fMRI experiments with different classification goals.

## 5.1. Algorithm choosing determination

Our toolbox extends the traditional network construction (PC-based FC network) to high-order FC networks and SR-based networks. It brings a new requirement to choose proper network construction methods as previous toolboxes did not provide such options. While in the toolbox we separate these network construction algorithms into two categories based on the requirement of parameter optimization, a more direct, operational based subdivision method, we could also separate them into pairwise (PC, tHOFC, aHOFC, and dHOFC) and multi-regional relationship-based methods (all SR-based algorithms). While the test-retest reliability of the PC and all HOFC algorithms has been verified (Zhang et al., 2017a), a similar study has not been done yet for SR-based network construction. Therefore, if users want to use a simple yet reliable network construction method, PC, tHOFC, and aHOFC are suggested. Compared to PC, tHOFC and aHOFC is more robust to noise yet interpretable, and most importantly, they could provide supplementary information to PC (Zhang et al., 2019a, 2016a). Therefore, they are suggested in such a scenario. From another viewpoint, only dHOFC utilized dynamic FC, while all other methods still focus on static FC. Collectively, all of the above general points could be put into consideration when choosing algorithms.

While featuring abundant network construction methods, the aim of such type of research could usually be making the classification as accurate as possible. However, we advise refraining from the urge of trying every method and report the best one, because choosing method itself can be a "parameter" optimization process, where the testing data should not be seen. Therefore, we try to provide some suggestions for users to choose a specific method, while it should always keep in mind that 1) not all the methods lead to good results, 2) each method has its preferences to specific data and research question, and 3) there is no absolute DO or DO NOT but previous studies have shown certain preferences. For example, it has been shown that there are fewer differences in PC-based static FC network comparison between patients with mental disorders and normal controls (Zheng et al., 2019). Instead, dynamic FC could reveal more group differences in such cases (Demirtaş et al., 2016; Kaiser et al., 2016; Rashid et al., 2014). By using the dynamic FC, dHOFC could capture more high-level complex interaction among brain regions and perform better than conventional low-order static FC approaches (Zheng et al., 2019). Therefore, for the disease that is assumed to have little alterations in the brain network, dHOFC is suggested.

Since most biological networks are intrinsically sparse (Rubinov and Sporns, 2010), SR has been widely used in biological signal analysis, such as electroencephalography (EEG) (Wen et al., 2016; Zhang et al., 2016b) and RS-fMRI (Lv et al., 2015; Suk et al., 2015). Another advantage of SR-based brain network construction with RS-fMRI is that the weak FC that could be induced by noise and artifacts can be suppressed to zero without an arbitrarily defined threshold (Wee et al., 2014). Therefore, for the data with the poten-tially higher noise level, SR-based methods can be used. The parameter(s) associated with this type of methods should be tuned to achieve better classification performance, which requires more data for the nested cross-validation. Therefore, SR-based methods, especially those with more freely estimable parameters, should be used with a large sample size. If users have a concern about robustness and parameter sensitivity, usually the issues associated in the SR-based analysis, they can still choose these methods, as the toolbox provides explicit and interpretable robustness and parameter sensitivity tests (see Sections 2.3.4 and 2.3.5). Since different SR-based methods incorporate different constraint into network modeling, users are advised to check Section 2.1.2 for specific benefit that each method can provide. For example, if the brain networks generated in Section 2.3.2 seem too sparse, the user may choose WSR, WSGR, or SLR to make the estimated network less sparse, contain more strong connections, or have certain structures (Qiao et al., 2016; Yu et al., 2017). If the data looks quite heterogeneous across subjects and the PC-based networks show large variability that might be caused by the noise and artifacts, users may choose group-wise sparse representation, such as GSR or SSGSR, to make the networks more topologically identical across individuals (Zhang et al., 2019c).

## 5.2. Comparison with other similar toolboxes

Before our toolbox, there exist several freely available packages for machine learning modeling of neuroimaging data, including PyMVPA, Sci-kit Learn, PRoNTo and GraphVar. We thus briefly compare them with our toolbox, as summarized in Table 6. PyMVPA (Hanke et al., 2009) and Scikit Learn (Abraham et al., 2014) are the sophisticated and flexible software packages primarily written in Python. The wide application of these two packages allows them to easily combine with a range of other neuroimaging and machine learning packages, including applications to magnetoencephalography (MEG), EEG, structural MRI and fMRI (Abraham et al., 2014; Guntupalli et al., 2018; Hanke et al., 2009). However, the two packages only provide command line-based analysis without any GUI, which is not easy to use by users without coding and Python knowledge. PRoNTo and GraphVar are both MATLAB toolboxes with GUI and provide abundant functions, including pattern recognition analysis for the analysis of neuroimaging data. PRoNTo aims at providing a comprehensive and user-friendly framework for multivariate analysis of neuroimaging data (Schrouff et al., 2013). GraphVar provides machine learning-based model construction, validation, and exploration, which can use graph measures (extracted from brain networks) to model their relationship with other variables, thus flexible for neuroimaging applications (Kruschwitz et al., 2015; Waller et al., 2018).

Compared to our toolbox, these toolboxes did not provide plenty (if any) of the state-of-the-art network construction methods such as the SR-based algorithms and the high-order network construction methods. We think that the network construction is as equal as, if not more important than, other processes such as feature reduction and classification, because it is the network features that are extracted for classification. Therefore, more effort should be put on developing brain network construction algorithms, as this plays an important role in the subsequent network-based classification studies.

**Table 5**
Comparison of the main features of the available software packages

|  | PyMVPA | Scikit-learn | PRoNTo | GraphVar | BrainNetClass |
|---|---|---|---|---|---|
| **Inputs** | Numpy arrays, *.txt, NIFTI, EEP | Numpy arrays, metadata | MRI/fMRI feature maps (NIFTI) | Time series, connectivity matrix (*.mat) | Time series (*.txt) |
| **Language** | Python | Python | Matlab | Matlab | Matlab |
| **Voxel/network-wise** | Voxel | Voxel/network | Voxel | Network | Network |
| **Interface** | Command Line | Command Line | GUI,batch, Command line | GUI | GUI,batch |
| **Static or dynamic FC** | Static | Static | Static | Static | Both |
| **Result display** | ✗ | ✗ | ✓ | ✓ | ✓ |
| **Network construction** | ✗ | ✗ | ✗ | ✓ | ✓ |
| **High-order FC or SR** | ✗ | ✗ | ✗ | ✗ | ✓ |
| **Contributing features** | ✗ | ✗ | ✗ | ✗ | ✓ |

Compared to these toolboxes, BrainNetClass provides comprehensive functions for result display, classification model evaluation, and interpretation (Section 2.3). Many practical features that have been reported in the previous disease classification studies are included, such as saving important features for visualization and constructed networks visualization (Chen et al., 2017a), as well as parameter sensitivity test (Yu et al., 2017; Zhang et al., 2019c). Some features that are essential for clinical applications such as model robustness test were not even reported in the previous studies. The above mentioned features are all provided by our toolbox.

### 5.3. Compatibility and computational requirement

The toolbox works well with Matlab 2016a and all later versions on a Linux desktop computer or Windows Personal Computer or computing servers. For the study involving a very large sample size, we recommend using a computer with large physical memory (e.g., at least 32 GB for 200 samples or larger for more samples). For the methods requiring no parameters, it will not take too long to finish the whole process. For methods requiring parameter tuning, BrainNetClass can distribute network constructions with different combinations of the parameter(s) to different CPU cores using the MATLAB parallel computing modules to save computational time. We also provide 10-fold cross validation if the study involves a larger sample size, as it will significantly decrease the computational time.

### 5.4. Limitations and future works

We provide BrainNetClass v1.0 to meet the urgent call for standardization of the methodology with an emphasis on (re-evaluation of) the reproducibility, generalizability, and interpretability of the (existing) network-based classification. It is yet in its first official version and unavoidably has limitations. First, it now only allows users to conduct two-class classification. We will add support vector regression (SVR) and multi-class classification in the future (Cui and Gong, 2018). Second, more feature extraction options will be provided beyond the current two types of features (FC links and local clustering coefficients) to better characterize network topology and further boost classification performance. Third, more sophisticated feature selection methods, such as ElasticNet and SVM wrapped method (recursive feature elimination, or SVM-RFE (Duan et al., 2005)), might be implemented in the future. Fourth, more options of classifiers, such as random forest and Naïve Bayes, could be provided, and the final classification result may come from the ensemble of multiple classifiers to improve performance. Fifth, more complex network definitions (e.g., hyper-connectivity-based network (Jie et al., 2016) and more dynamic FC-based network construction methods (e.g., the variability of the dynamic FC (Chen et al., 2017b)) should be added. Finally, the optimization of parameters for network construction and classification could be the most important yet difficult problem in the current studies. We only allow at most two freely estimable parameters at the current stage to compromise between the computational timing/memory required and the modeling flexibility. In the future, with well-designed parameter optimization strategies (e.g., adaptive parameter range determination), we might allow more parameters to be simultaneously optimized.

## 6. Conclusions

We introduce a novel, Matlab GUI-based, open-coded, fully automated brain functional network construction and classification toolbox, namely BrainNetClass v1.0. It allows users to construct brain networks with advanced methods and conduct rigorous feature extraction/selection, parameter optimization, classification, and model evaluation in a standard and well-accepted framework. This toolbox is the first one that is friendly to neuroscientists and clinicians for facilitating their function connectomics-based disease diagnosis or classifications with comprehensive yet intuitive and interpretable results. It is helpful to standard-ize the methodology and boost the clinical application of neuroimaging-based machine learning with improved reproducibility, generalizability, and interpretability. The toolbox, manual, and exemplary datasets are available at https://github.com/zzstefan/BrainNetClass.

## Acknowledgements

HZ and DS were supported in part by NIH grants (EB022880, AG049371, AG042599, and AG041721). XC, YZ, LQ, and RY were supported in part by NIH grants (EB022880).